\documentclass{iaus}
\usepackage{graphicx}

\title[Leptonic emission from microquasar jets] %% give here short title %%
{Leptonic emission from microquasar jets: from radio to very high-energy gamma-rays}

\author[Bosch-Ramon, V. et al.]   %% give here short author list %%
{Valent\'i Bosch-Ramon$^1$, Josep M. Paredes$^1$ \and
Gustavo E. Romero$^{2,3}$}

\affiliation{$^1$Departament d'Astronomia i Meteorologia, 
Universitat de Barcelona, Av. Diagonal 647, E-08028 Barcelona, 
Catalonia (Spain), email: vbosch@am.ub.es \\[\affilskip]
$^2$Instituto Argentino de Radioastronom\'ia, C.C.5, (1894) Villa Elisa, 
Buenos Aires (Argentina) \\[\affilskip]
$^3$Facultad de Ciencias Astron\'omicas y Geof\'isicas, UNLP,  
Paseo del Bosque, 1900 La Plata (Argentina)}

\pubyear{2005}
\volume{230}  %% insert here IAU Symposium No.
\pagerange{119--126}
\date{?? and in revised form ??}
\jname{Proceedings Title IAU Symposium}
\editors{A.C. Editor, B.D. Editor \& C.E. Editor, eds.}
\begin{document}

\maketitle

\begin{abstract}
Microquasars are sources of very high-energy gamma-rays and, very probably,
high-energy gamma-ray emitters. We propose a model for a jet that can
allow to give accurate observational predictions for jet emission at
different energies and provide
with physical information of the object using multiwavelength data.
\end{abstract}

%\section{Introduction}

Microquasars are X-ray binaries with relativistic jets whose emission extends
from radio to gamma-rays. In particular, LS 5039, discovered by Paredes et al.
(2000), turned out to be the first likely high-energy gamma-ray microquasar due to its
possible association with the EGRET source 3EG~J1824$-$1514 (Paredes et al.
2000). Further theoretical studies furnished with more reliability this
association (Bosch-Ramon \& Paredes 2004), which could be extended to other
EGRET sources (Kaufman Bernad\'o et al. 2002, Romero et al. 2003, Bosch-Ramon
et al. 2005a). Very recently, Aharonian et al. (2005) have published the
detection of the microquasar LS 5039 at TeV energies, leaving no doubt about
the gamma-ray emitting nature of these objects. The aim of the present work is
to show that from the theoretical point of view microquasar jets can be sources 
of emission with energies covering the whole spectral band.

In our model, the jet is modelled as dynamically dominated by cold protons and
radiatively dominated by relativistic leptons. The characteristics of the orbit 
and the companion star constrain the way in which the accretion takes place. This 
has been taken into accout for a consistent orbital variability treatment. 
The magnetic field energy
density and the non-thermal particle maximum energy values along the jet depend
on the cold matter energy density and the particle acceleration/energy loss
balance, respectively, and the amount of relativistic particles within the jet
is restricted by the efficiency of the shock to transfer energy to them. The
model takes into account the external and internal photon and matter
fields, which interact with relativistic particles in the magnetized jet,
producing emission from radio to very high energies. Concerning the 
the physical parameters of the model, we have adopted the typical values
for a massive X-ray binary system and the standard ones in jet and acceleration theory
(OB star, semi-major axis of about 0.1 AU, strong stellar wind, jet
matter/accretion matter rate ratio of about 0.1, magnetic field of about a 
10\% of the equipartition one, acceleration efficiencies 10$^{-3}$-10$^{-1}$qBc). For
further details concerning the model, see Bosch-Ramon et al. (2005b).

This model has been applied to LS 5039. In Fig. 1, we show the
computed SED, and the flux and photon index evolution along the orbit for
this source. The model underpredicts the fluxes in the radio and TeV band.
Although the difference is less than one order of magnitude, it could be
interpreted as a hint that more intense high energy processes (particle
acceleration and emission) are taking place already outside the binary system,
since radiation is emitted mainly within the 100 GeV-photon absorption region
close to the companion star. This could also lead to higher fluxes in radio,
due to this radiation is optically thin when emitted outside the binary system.
Concerning the variability of the source along the orbit, we have used a
particular accretion model for this source consisting in a slow equatorial wind
from a fast rotating stellar companion (Casares et al. 2005), altogether with a
stream of matter ejected during the periastron passage and reaching the compact
object at phase 0.8--0.9, when the X-ray (Bosch-Ramon et al. 2005c) and the TeV
peaks (Aharonian et al. 2005, Casares et al. 2005) are observed. For more
details concerning the application of the model to LS 5039, see the work of
Paredes et al. (2005). It is worth noting that the total jet kinetic power
required for producing a SED like the one plotted in Fig.~3 is few times 10$^{36}$
erg/s. Therefore, the very likely nature of microquasars and, in particular, LS
5039 as GeV-TeV emitters appears well founded from theoretical grounds. 

\begin{figure}
\includegraphics[height=2in,width=5.3in,angle=0]{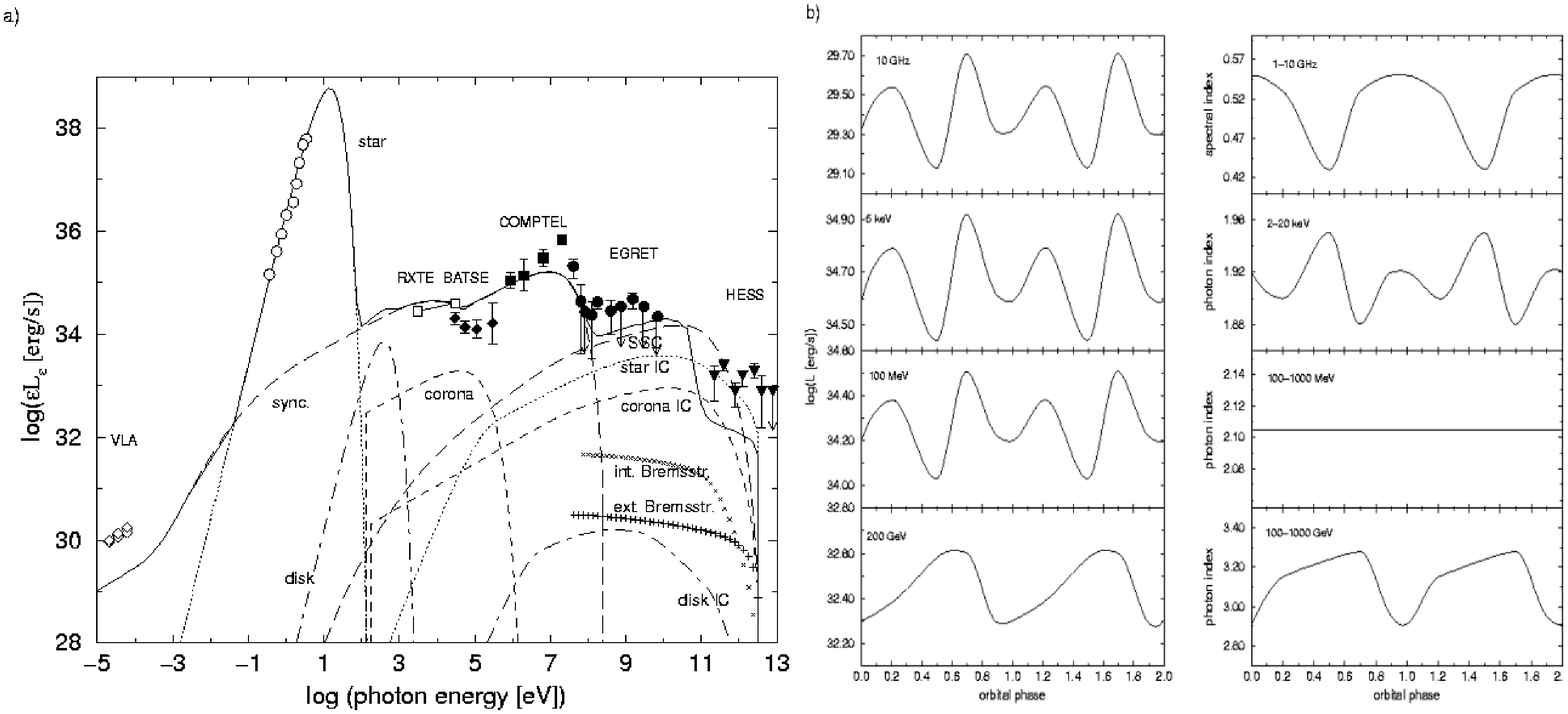}
  \caption{a) Computed SED for LS 5039 and b) its flux 
and photon index evolution at different energy bands.}\label{}
\end{figure}

\begin{acknowledgments}
V.B-R. and J.M.P. acknowledge partial support by DGI of the spanish Ministerio
de Educaci\'on y Ciencia under grant AYA2004-07171-C02-01, as well as additional
support from the European Regional Development Fund  (ERDF/FEDER). During this
work, V.B-R. has been supported by the DGI of the spanish Ministerio de
Educaci\'on y Ciencia under the fellowship BES-2002-2699. G.E.R is supported by
the Argentine Agencies CONICET and ANPCyT (PICT 03-13291).
\end{acknowledgments}

\end{document}